\documentstyle[11pt]{article}

\def\hybrid{\topmargin -20pt  \oddsidemargin 0pt
      \headheight 0pt   \headsep 0pt
      \textwidth 6.25in % A4 paper
      \textheight 9.5in % A4 paper
      \marginparwidth .875in
      \parskip 5pt plus 1pt   \jot = 1.5ex}

\hybrid

\def\x{\times}

\def\o+{\oplus}

\def\lra{\longrightarrow}

\def\beqa{\begin{eqnarray}}
\def\eeqa{\end{eqnarray}}
                       
\newcommand{\si}{\sigma}

\newcommand{\om}{\omega}

\begin{document}
\thispagestyle{empty}
\rightline{LMU-ASC {104}/10}
\vspace{2truecm}
\centerline{\bf \LARGE Spectral Bundles and the DRY-Conjecture}

\vspace{1.5truecm}
\centerline{Bj\"orn Andreas\footnote{andreas@math.hu-berlin.de}$^\$$ 
and Gottfried Curio\footnote{gottfried.curio@physik.uni-muenchen.de; 
supported by DFG grant CU 191/1-1}$^{\scriptsize\mbox{\pounds}}$} 

\vspace{.6truecm}

\centerline{$^\$${\em Institut f\"ur Mathematik}}
\centerline{{\em Humboldt-Universit\"at zu Berlin}}
\centerline{{\em Rudower Chaussee 25, 12489 Berlin, Germany}}

\vspace{.3truecm}
\centerline{$^{\scriptsize\mbox{\pounds}}${\em Arnold-Sommerfeld-Center 
for Theoretical Physics}}
\centerline{{\em Department f\"ur Physik, 
Ludwig-Maximilians-Universit\"at M\"unchen}}
\centerline{{\em Theresienstr. 37, 80333 M\"unchen, Germany}}

\vspace{1.0truecm}

\begin{abstract}
Supersymmetric heterotic string models, built from a Calabi-Yau threefold $X$ 
endowed with a stable vector bundle $V$, usually start from a
phenomenologically motivated choice of a bundle $V_v$ in the visible sector,
the spectral cover construction on an elliptically fibered $X$
being a prominent example.
The ensuing anomaly mismatch between $c_2(V_v)$ and $c_2(X)$, or rather
the corresponding differential forms,
is often 'solved', on the cohomological level, by including a fivebrane.
This leads to the question whether the 
difference can be alternatively realized by a further stable bundle.
The 'DRY'-conjecture of Douglas, Reinbacher and Yau in math.AG/0604597
gives a sufficient condition on cohomology classes on $X$ to be realized
as the Chern classes of a stable sheaf. In 1010.1644 [hep-th] we showed
that infinitely many classes on $X$ exist for which the conjecture ist true.
In this note we give the sufficient condition for the mentioned 
fivebrane classes to be realized by a further stable bundle in the 
hidden sector.
Using a result obtained in 1011.6246 [hep-th] we show that corresponding
bundles exist, thereby confirming this version of the DRY-Conjecture.
\end{abstract}

\newpage

To get fourdimensional $N=1$ supersymmetric models 
from the tendimensional $E_8\x E_8$ heterotic string 
one compactifies on a Calabi-Yau threefold $X$ endowed with a polystable 
holomorphic vector bundle $V'=(V_v, V_{h})$. 
$V_v$ is usually a stable bundle considered to be embedded in (the visible) $E_8$ 
($V_{h}$ plays the corresponding role for the hidden $E_8$).

Usually one specifies $V_v$ to fulfill some phenomenological requirements, 
like the generation number. This specifies discrete parameters of the
bundle construction and anomaly freedom of the construction is encoded
in the integrability condition for the existence 
of a solution to the anomaly cancellation equation
\beqa
\label{anom}
c_2(X)&=&c_2(V_v)+W.
\eeqa
Here $W$, as it stands, has just the meaning to indicate a possible mismatch
for a certain bundle $V_v$; it can be understood either as the cohomology class
of (the compact part of the world-volume of) a fivebrane, or as second Chern
class of a further stable bundle $V_{h}$ in the hidden sector. In
the first case the class of $W$ has to be effective for supersymmetry
to be preserved. 
If one wants to solve without a fivebrane one
has to make sure that a corresponding hidden bundle $V_h$ with
having $c_2(V_h)=W:=c_2(X)-c_2(V_v)$ exists.

In [\ref{AGF1}] it has been shown that when (\ref{anom})
can be satisfied with $W=0$ then $X$ and $V_v$ can be 
deformed to a solution of the anomaly equation even already on the level 
of differential forms. 

This leads to the general question to give sufficient conditions for the
existence of stable bundles with prescribed Chern class $c_2(V)$.
Concerning this a conjecture is put forward
in [\ref{DRY}] by Douglas, Reinbacher and Yau (DRY) (actually we use the 
particular case of the conjecture with $c_1(V)=0$). We will 
actually use a weaker version of the conjecture, considered already 
in [{\ref{DRYAC}]. We recall the following definition and conjecture.

{\bf Definition.} Let $X$ be a Calabi-Yau threefold of $\pi_1(X)=0$ and
$c\in H^4(X, {\bf Z})$, \\
i) $c$ is called a {\em Chern class} 
if a stable $SU({\cal N})$ vector bundle $V$ on $X$ exists with $c=c_2(V)$\\
ii) $c$ is called a {\em DRY class} if an ample class 
$H\in H^2(X, {\bf R})$ exists (and an integer ${\cal N}$) with 
\beqa
\label{weak dry}
c_2(V)&=&{\cal N}\Bigg( H^2+\frac{c_2(X)}{24}\Bigg).
\eeqa

{\bf Weak DRY-Conjecture.} {\em On a Calabi-Yau threefold $X$ of $\pi_1(X)=0$ 
every DRY class $c\in H^4(X, {\bf Z})$ is a Chern class.}

Here it is understood that the integer ${\cal N}$ occurring in the two
definitions is the same.

We choose $X$ to be elliptically fibered over a rational base surface $B$
with section $\sigma:B\lra X$ (we will also denote by $\si$ 
the embedded subvariety
$\si(B)\subset X$ and its cohomology class in $H^2(X, {\bf Z})$), 
a case particularly well studied in phenomenological applications. 
Typical examples for $B$ are Hirzebruch or 
del Pezzo surfaces (or suitable blow-ups of these).
As in [\ref{DRYAC}}] we consider bases $B$ for which $c_1:=c_1(B)$ is ample
(this excludes the Hirzebruch surface ${\bf F_2}$).
(The classes $c_1^2$ and $c_2:=c_2(B)$ will be considered as (integral) 
numbers.)

On $X$ one has according to the general 
decomposition
$H^4(X, {\bf Z})\cong H^2(X, {\bf Z})\si \oplus H^4(B, {\bf Z})$
\beqa
c_2(V)&=&\phi \si + \om
\eeqa
where $\om$ is understood as an integral number 
(pullbacks from $B$ to $X$ will be usually suppressed).
Similarly one has $c_2(X)=12c_1\si +c_2+11c_1^2$ or, with Noethers theorem, 
$12 c_1 \si +10c_1^2+12$.

One gets the following theorem [\ref{DRYAC}]

\noindent
{\bf Theorem on DRY classes.} 
{\em A class $c=\phi\si + \omega \in H^4(X, {\bf Z})$ is a DRY class if and 
only if the following condition is fulfilled
(where $b$ is some $b\in{\bf R}^{>0}$ and 
$\omega \in H^4(B, {\bf Z})\cong {\bf Z}$):\\
$\phi-{\cal N}(\frac{1}{2}+b)c_1$ is ample and 
$\frac{1}{\cal N}\omega> \omega_0(\phi; b):=r+\frac{c_1^2}{4}(b+\frac{q}{b})$}.

Here one uses the following abbreviations, cf.~[\ref{DRYAC}]
\beqa
r:=\frac{1}{2{\cal N}}\phi c_1+\frac{1}{6}c_1^2+\frac{1}{2}\\
q:=\frac{(\phi - \frac{{\cal N}}{2}c_1)^2}{{\cal N}^2c_1^2}.
\eeqa
Note that under the hypthesis that 
$\phi - \frac{{\cal N}}{2}c_1=A+b{\cal N}c_1$ with an ample class $A$ 
on $B$ (and with $c_1\neq 0$ effective) one has 
$(\phi - \frac{{\cal N}}{2}c_1)^2>b^2{\cal N}^2c_1^2$,
thus one has $b<\sqrt{q}$.

To make the condition more explicit let us choose a concrete $b$: 
the choice $b=1/2$, for example, gives the conditions
\beqa
\phi - {\cal N}c_1 &\;\; & \mbox{ample}\\
\frac{1}{\cal N}\omega \;\;\;\; 
& > & \frac{5}{12}c_1^2 + \frac{1}{2} + \frac{1}{2{\cal N}^2}\phi^2.
\eeqa

These are the conditions for the application of the weak DRY conjecture.
By contrast recall that the starting conditions of the effectivity of 
the fivebrane class $W=c$ were just
\beqa
\phi & \;\; & \mbox{effective}\\
\omega & \geq & 0.
\eeqa

In the case of our application, where $c$ arises as $c_2(X)-c_2(V_v)$
for an $SU(n)$ 
spectral cover bundle $V_v$, one has for the effective fivebrane class
$W=c$ the following
\beqa
\phi   &=& 12c_1 - \eta_v \\
\label{omega 5brane}
\omega &=& 10c_1^2+12 
+\frac{n^3-n}{24}c_1^2-(\lambda_v^2-\frac{1}{4})\frac{n}{2}\eta_v(\eta_v-nc_1).
\eeqa

Here $\eta_v$ is an effective class in $B$ with $\eta_v-nc_1$ 
also effective and $\lambda_v$
is a half-integer satisfying the following conditions: $\lambda_v$ 
is strictly half-integral for $n$ being odd;
for $n$ even an integral $\lambda_v$ requires 
$\eta_v\equiv c_1 \  ({\rm mod}\  2)$ while a strictly 
half-integral $\lambda_v$ requires $c_1$ even. (Often one assumes 
that $\eta_v-nc_1$ is not only effective but even ample in $B$.) 

As the visible bundle is a spectral cover bundle, and so is stable with respect
to the typical spectral K\"ahler class (the polarization) 
$H=\epsilon H_0 +H_B$ (where $H_0$ is an ample class on $X$, $\epsilon$ is 
chosen sufficiently small [\ref{FMW3}] and $H_B$ is an ample class on $B$)
we have to choose a hidden bundle $V_h$ which is stable, or polystable, with
respect to the same class. This suggests to take for $V_h$ also a spectral
bundle (which we take with ample spectral cover surface). 
This will allow to produce in $c_2(V_h)$ the part $\phi\si$ of
the needed class $W=c=\phi\si+\omega$. However the special form of the
remaining fiber term of $c_2(V_h)$ for a spectral bundle $V_h$ will usually
not be able to represent a given $\omega$ part in $c$.
Therefore one needs a second input to match also this part of $c$.
Suitable to represent this part is a pull-back bundle $\pi^* E$
(where $E$ is a stable bundle on $B$) as it has $c_2(\pi^* E)=k$, 
and so consists just of the needed part where the integer $k$ is arbitrarily
choosable as long $k\geq r(E)+2$ where $r(E):=rank(E)$ 
[\ref{Art}].
Therefore, in total, we will choose in the hidden sector a combination
of a spectral $SU(N)$ bundle and a pull-back bundle $\pi^* E$ of rank $r(E)$
such that ${\cal N}=N+r(E)$.
It is possible to combine the advantages of a hidden 
spectral bundle with those of a hidden pullback bundle because,
according to [\ref{AGF2}], the pullback bundle will be stable with respect to 
the same polarization as the spectral bundle.

Thus combining the flexibilities of the (hidden) spectral bundle 
in the $\phi$-term (in the $c=\phi \si + \omega$ decomposition of 
$c_2(V_h)+c_2(\pi^*E)$; we call this the ``$\si$''-term of $c_2$) 
with that of the the pullback bundle in 
the $\omega$-term (which we call the ``fiber''-term)
we will be able, as shown below, 
to represent a class $c=\phi \si + \omega$
as $c_2$ of the total hidden (sum) bundle if it satisfies the following
conditions (referred to below as conditions of ``realizability'' of the class
$c$)
\beqa
\label{Chern condition1}
\phi -N c_1 &\;\;& \mbox{ample} \\
\label{Chern condition2}
\omega &\geq & 1.
\eeqa

On the other hand, the conditions for a DRY class were
\beqa
\label{DRY condition1}
\phi-{\cal N}\Big(\frac{1}{2}+b\Big)c_1 &\;\; & \mbox{ample}\\
\label{DRY condition2}
\frac{1}{{\cal N}}\omega & > & r+\frac{c_1^2}{4}\Big(b+\frac{q}{b}\Big).
\eeqa

Thus one sees that for $N=2=r(E)$ all DRY classes are realised as
second Chern classes of a polystable bundle $V_h \oplus \pi^*E$:
for a DRY class one has that $\phi-\frac{{\cal N}}{2}c_1$ is ample, 
so the condition in
(\ref{Chern condition1}) is satisfied (this works always as long as one has
${\cal N}/2\geq N$, i.e. $r(E)\geq N$). Furthermore the necessary 
condition $\frac{1}{{\cal N}}\omega\geq r+s$ for the fiber part of a DRY 
class (cf.~[\ref{DRYAC}], here $s=\frac{1}{2{\cal N}}
\sqrt{c_1^2}\sqrt{(\phi-\frac{{\cal N}}{2}c_1)^2}$) implies that 
\beqa
\omega &\geq & \frac{1}{2}\phi c_1 + \frac{{\cal N}}{6}c_1^2 
+ \frac{{\cal N}}{2}
+\frac{1}{2}\sqrt{c_1^2}\sqrt{(\phi-\frac{{\cal N}}{2}c_1)^2}.
\eeqa
Thus this gives for ${\cal N}=4$ that one has (\ref{Chern condition2})
for the fiber part of all DRY classes.

{\em Thus we have shown the weak DRY conjecture for the rank $4$ case}, in the
sense of realising a DRY class by the $c_2$ of a corresponding rank $4$
bundle; however we have shown only a realisation by a polystable bundle
which is a sum $V_h\oplus \pi^*E$. (Starting from this bundle one can build,
as a variant of this construction,
an irreducible, stable bundle, constructed as a non-split extension.)

It remains to check that all classes, as described in (\ref{Chern condition1})
and (\ref{Chern condition2}), are indeed realizable a $c_2(V_h)+c_2(\pi^*E)$.
First notice that one has
\beqa
c_2(V_h\oplus \pi^*E)&=&\eta_h \si
-\frac{N^3-N}{24}c_1^2+\frac{N}{2}(\lambda_h^2-\frac{1}{4})\eta_h(\eta_h-Nc_1)
+k.
\eeqa
The $\si$ term is clear, one takes $\eta_h:=\phi$.
The instanton number $k$ will be arbitrarily specifiable [\ref{Art}]
if one has $k\geq rk(E)+2$.
For the special case $r(E)=2$ one has, however, the stronger result
that $k\geq 2$ is sufficient, cf.~[\ref{Fr}], Ch.~10, Th.~37.
So, to have arbitrary $\omega\geq 1$ realizable, 
we need to produce a $V_h$ with the fiber term of $c_2(V_h)$ being $\leq -1$.
Having $N=2$ and taking the optimal possibility $\lambda_h=0$
we get for this fiber part 
$-\frac{1}{4}c_1^2-\frac{1}{4}\eta_h(\eta_h-2c_1)
=-\frac{1}{4}(\eta - c_1)^2$;
furthermore the conditions in the spectral construction demand in this case
that $\eta_h\equiv c_1 (\mbox{mod} \; 2)$, i.e.,~$\eta_h=c_1+2a$ 
(for an integral class $a$, which is nonzero as $\eta-2c_1$ has to 
be effective), thus leaving the term $-a^2$.
Now one has that actually $a^2>0$:
for one has $c_1^2>0$ and $\eta_h(\eta_h-2c_1)> 0$ 
because $\eta_h-2c_1$ and $c_1$ are here ample.

Often one will have the sharper $a^2\geq 2$ which will allow $\omega \geq 0$
in (\ref{Chern condition2}): if $c_1$ is even, for example
(because of $a^2\equiv a c_1\; (\mbox{mod}\, 2)$), or on a Hirzebruch surface
(where already $\frac{1}{4}c_1^2=2$).

Finally let us come back to the original problem which motivated our
consideration of the DRY conjecture.
Notice that, after our consideration of the question which classes $c$
are realizable as Chern classes, the focus of the original question has actually
somewhat shifted: it will be enough to investigate, whether the 
fivebrane class $W$ belongs to these ``realizable'' classes; for note
that in any case the DRY class condition is only sufficient for a class
to be realizable as Chern class.

This demand of ``realizability'' means, 
concerning the $\si$ part, 
whether $\phi=12c_1-\eta_v$, which is effective by assumption, fulfills
actually the stronger condition that $\phi-2c_1$ is ample; and
concerning the fiber part, whether 
$\omega$ in (\ref{omega 5brane}), which by assumption is $\geq 0$, is
actually $>0$. 

Let us conclude: usually one supplemented a phenomenolgical
heterotic spectral bundle construction, 
which was constructed in the visible sector, 
with a fivebrane class $W$
(to solve the anomaly cancellation equation); then
the only condition one had to satisfy was that this class 
$W=c=\phi\si + \omega$ is effective (i.e. $\phi$ effective and $\omega\geq 0$).
Here we have seen that, sharpening this demand just slightly 
(to $\phi-2c_1$ ample and $\omega > 0$) 
one can actually turn on a polystable bundle $V_h\oplus \pi^*E$
in the hidden sector whose $c_2$ just realizes the class $W$
(furthermore the total bundle including also $V_v$ is polystable).
One can thus solve the anomaly for (almost) all the visible spectral
bundles (for which one can solve it {\em with} fivebrane)
also {\em without} any fivebrane (and instead just with a hidden bundle).
Then one can actually solve the anomaly already on the level of differential forms [\ref{AGF2}].

\section*{References}
\begin{enumerate}

\item
\label{DRY}
M.R.~Douglas, R.~Reinbacher and S.-T.~Yau,
{\em Branes, Bundles and Attractors: Bogomolov and Beyond},
math.AG/0604597.

\item
\label{FMW}
R. Friedman, J. Morgan and E. Witten, {\em Vector Bundles and F-Theory},
hep-th/9701162, Comm. Math. Phys. {\bf 187} (1997) 679.

\item
\label{FMW3}
R. Friedman, J. Morgan and E. Witten, 
{\em Vector Bundles over Elliptic Fibrations}, arXiv:alg-geom/9709029.

\item
\label{AC1}
B. Andreas and G. Curio, {\em Stable Bundle Extensions on elliptic 
Calabi-Yau threefold},
J. Geom. Phys. {\bf 57}, 2249-2262, 2007, math.AG/0611762.

\item
\label{AGF1}
B. Andreas and M. Garcia-Fernandez, {\em Solution of the Strominger 
System via Stable Bundles on Calabi-Yau threefolds}, arXiv:1008.1018 [math.DG].

\item
\label{AGF2}B. Andreas and M. Garcia-Fernandez, {\em Heterotic Non-K\"ahler
Geometries via Polystable Bundles on Calabi-Yau Threefolds}, arXiv:1011.6246.

\item
\label{DRYAC}
B. Andreas and G. Curio, {\em On possible Chern Classes of stable Bundles on
Calabi-Yau threefolds}, arXiv:1010.1644 [hep-th].

\item
\label{Art}
I.V.~Artamkin, {\em Deforming Torsion-free Sheaves on an Algebraic Surface},
Math.USSR.Izv. {\bf 36} (1991) 449.

\item
\label{Fr}
R.~Friedman, {\em Algebraic Surfaces and Holomorphic Vector Bundles},
Springer (1998).

\end{enumerate}
\end{document}